\documentclass[12pt]{article}
\usepackage{graphicx}
\usepackage{amssymb}

\def\vv{\vspace{-1ex}}

\begin{document}

\begin{center}

{\Large
Use of Silicon Photomultipliers in ZnS:$^6$LiF scintillation neutron detectors:
signal extraction in presence of high dark count rates
}

\vspace{2ex}
A.\,Stoykov, J.-B.\,Mosset, U.\,Greuter, M.\,Hildebrandt, N.\,Schlumpf

\vspace{2ex}

Paul Scherrer Institut, CH-5232 Villigen PSI, Switzerland \\[1ex]

\end{center}

\vspace{2ex}
\noindent
We report on the possibility of using Silicon Photomultipliers (SiPMs)
to detect the scintillation light from neutron conversion in ZnS:$^6$LiF scintillators.
The light is collected by wavelength-shifting fibers embedded into the scintillator.
The difficulty of extracting neutron signals in the presence of high dark count
rates of the SiPMs is addressed by applying a dedicated processing algorithm to analyze
the temporal distribution of the SiPM pulses.
With a single-channel prototype detection unit we demonstrate a very good
neutron signal extraction at SiPM dark count rates of about 1\,MHz.

\vspace{3ex}

\vspace{2ex}
\section{Introduction}
The current approaches \cite{Rhodes97,Sakasai09,Nakamura12,Nakamura14}
for large-area multi-channel detectors
for thermal neutrons which are based on the scintillation process
in ZnS:$^6$LiF (ZnS:$^{10}$B$_2$0$_3$) screens use photomultiplier tubes (PMTs)
to detect the scintillation light collected through clear or wavelength-shifting (WLS) fibers.
Equipping each detection channel with its own single-anode PMT is not realistic
due to cost and detector volume considerations. Light-sharing schemes are applied
to reduce the total number of PMTs to be built in. Using multi-anode PMTs (MaPMTs) in combination
with WLS fibers individual channel readout becomes feasible.
Silicon Photomultipliers (SiPMs) \cite{Dolgoshein06},
being compact and significantly less expensive than MaPMTs,
could turn out to be advantageous if being used in readout schemes utilizing WLS fibers.
But, in contrast to MaPMTs, the intrinsic dark count rates of SiPMs
operated at room temperature are orders of magnitude higher.
Accordingly, a direct replacement of MaPMTs by SiPMs in the neutron detection systems
would require substantial reduction of the SiPM dark count rate which can only be achieved
through a deep cooling of the photosensor \cite{Dolgoshein06,Collazuol11}.
Although technically this approach is clearly feasible, the necessity to
cool the SiPMs would add to the complexity and to the cost of the detection system
which may outweigh all the advantages expected from their usage.

In this work we aim at developing a SiPM based ZnS:$^6$LiF neutron
detector operated at room temperature.
The long emission time of the ZnS scintillator, where only 25\,\% (60\,\%) of the photons
are emitted within the first 1\,$\mu$s (10\,$\mu$s) of a neutron scintillation event \cite{Kuzmin02},
and the deficient light collection due to the poor transparency of the scintillator
make it difficult to combine a high trigger efficiency for the neutron signals
with a reasonable suppression of the SiPM dark counts.
We solve this problem by optimizing the light collection from the scintillator
and by developing a signal processing system capable to reliably identify neutron signals
against the high background of the SiPM dark counts.

\section{Requirements on the prototype detector}
The main performance characteristics of ZnS:$^6$LiF (ZnS:$^{10}$B$_2$0$_3$) neutron detectors
are neutron detection efficiency, which is a product of the neutron absorption probability
in the scintillator and the trigger efficiency of the signal processing system,
background (quiet) count rate, gamma-sensitivity,
pulse-pair resolution (dead time), and multi-count ratio.
Typical values for these parameters can be found in \cite{Rhodes97,Sakasai09,Nakamura12,Nakamura14}.
It is interesting to note the following:
the trigger efficiency has negative correlation with all other parameters,
i.e. by lowering the efficiency the detector performance in all other aspects is improved.
With a low dark-count rate photosensor (PMT) it is in principle possible to have
at the same time a trigger efficiency close to 100\,\% and a low background count rate.
In practice, however,
it would require to operate the detector with a rather long dead time
to ensure that the system does not trigger on the afterglow photons
and the requirement on the multi-count ratio is also fulfilled.
Having the dead time in the order of 5\,$\mu$s
(typical value to ensure reasonable rate capabilities of the detector)
together with the requirement on the multicount ratio (typically below 1\,\%)
essentially leads to trigger efficiency values well below 100\,\%
(even down to $\sim 60$\,\% as in \cite{Nakamura12}).
With the low dark count rate photosensor (PMT) one faces, in principle,
the same problem, as with a high dark count rate photosensor (SiPM).
Namely in both cases one has to suppress unwanted counts. Only in the case of PMTs these
counts correspond to afterglow photons and in case of SiPMs to both afterglow photons and dark
counts.

With the single-channel prototype detector presented in this work we aim to satisfy the
requirements of the POLDI time-of-flight diffractometer \cite{Mosset13} on
the neutron detection efficiency ($\sim 65$\,\% at the wavelength of 1.2\,\AA) and
on the background count rate ($\sim 3\cdot10^{-3}$\,Hz). The required neutron detection efficiency
is to be achieved as a product of $\sim 80$\,\% neutron absorption probability in the scintillator
and the trigger efficiency of the signal processing system of $\sim 80$\,\%.
Evaluation of other detector characteristics as well as optimization studies
(which have to be performed taking into account all the detector parameters)
will be a subject of further investigations.

The above specified detector parameters have to be satisfied
at SiPM dark count rates up to $\sim 1$\,MHz.
This requirement comes from the following consideration.
Hamamatsu SiPMs of type MPPC S12571-025C or MPPC S12571-050C with 1\,x\,1\,mm$^2$ active area
are planned to be used.
At room temperature the dark count rate of these devices is $\sim 100$\,kHz \cite{Hamamatsu}.
During operation of the detection system we expect, taking into account the radiation conditions
at POLDI, an increase of the SiPM dark count rate by about 90\,kHz per year
conditioned by an increase of the concentration of radiation defects in silicon \cite{Musienko07}.
Accordingly, requiring that the detector parameters have
to be guaranteed up to $\sim 1$\,MHz dark count rate of the SiPM
is equivalent to the requirement that
they are to be guaranteed for about 10 years of detector operation.

\section{Test scintillator structure}
Figure~1 shows a cross-section of the single-channel scintillator unit (see also \cite{Mosset13})
used in the current tests. The unit consists of two layers (0.25\,mm and 0.45\,mm thick)
of ZnS:$^6$LiF scintillation material
(ND2:1 neutron detection screens from Applied Scintillation Technologies \cite{AST})
glued together using EJ-500 optical epoxy from Eljen \cite{Eljen}.
Four WLS fibers of type Y11(200)M from Kuraray \cite{Kuraray} are glued
(with the same epoxy) into the grooves made in the thicker layer.
At one side of the unit the fibers are cut along its edge and polished.
An aluminized Mylar foil serving as specular reflector is glued on these polished fiber ends
to increase the light yield at the other fiber ends connected to a SiPM.
The effective channel width is defined by the area covered by the fibers and is about 2.4\,mm.
The additional space to the total width of 7\,mm is used for handling purposes only.
The length of the structure is 50\,mm.

In a later multi-channel detector this 2.4\,mm wide groove/WLS fiber pattern
without the additional handling space will be repeated to cover the needed
neutron sensitive area. By stacking 4 scintillator units of this type,
one behind each other, the required neutron absorption probability
of $\sim 80$\,\% for 1.2\,\AA\ neutrons will be achieved.
In this estimate we use the value of $1.4\cdot10^{22}$~cm$^{-3}$
for the concentration of $^6$Li atoms in the ND2:1 screens
which is deduced from our own absorption measurements.

\section{Photon detection and signal processing}
In the present measurements a 3\,x\,3\,mm$^2$ MPPC S12572-050C from Hamamatsu \cite{Hamamatsu}
was used as the photon detector.
The SiPM was operated with a bias voltage of 67.8\,V (overvoltage 2.3\,V) at 23.5\,$^{\rm o}$C.
The dark count rate at these operating conditions is $\sim 1$\,MHz.

The implemented signal processing algorithm is not following the way of integrating
the photosensors analog signals with subsequent pulse-duration analysis like,
for example, in \cite{Kuzmin02}.
In the case of SiPMs, where fluctuations of one-electron signals can be rather large
as a result of the optical cross-talk, such integration would add to the fluctuations
of the resulting signal and thus deteriorate the dark-count rejection capability of the
detector.
The signal processing system proposed here is based on
a fast single photon detection followed by a time domain filtering.
The basic elements of the processing chain for the SiPM signals are presented in Figure~2.
It includes a high band-width amplifier, a leading-edge discriminator, a multistage filter,
and an event generator unit.

The SiPM signal is amplified and shaped by a wide band-width gain stage;
each SiPM pulse corresponds to one primary charge carrier triggering an avalanche breakdown
in one (or more than one by optical cross-talk) of the SiPM cells.
This signal is fed into a fast leading edge discrimination stage accepting all SiPM pulses
generated either by scintillation photons or thermally.
Figure~3 shows a scope screenshot (persistence mode) of the amplified SiPM pulses (SA)
and the generated discriminator signals (SD).
This kind of ``digitalization'' eliminates the effect of the SiPM cross-talk
on the performance of the detector and allows for a further signal processing scheme
being independent of the SiPM type.

The SD pulse sequence is processed by a ``single-pulse elimination'' filter implementation
(see Figure~4). The delay lines are chosen in order to prevent the initiating SD
respectively SF(i) pulses to pass the coincidence-AND gates.
To pass this type of filtering stage, a pulse at its input has therefore to show up
with a preceding pulse (being part of a pulse group) within the time interval
defined by the width of the internal gate (retriggerable mono-flop type) signal,
i.e. single pulses and always the first pulse from a group of consecutive pulses
are removed from the SD respectively SF(i) pulse sequence.
At the ED(i) outputs the number of remaining events
(pulse groups built by the retriggerable mono-flops) can be counted.
This counting is only done to analyse the filter performance,
e.g. to determine the remaining dark count rate dependence on the number of stages,
and is not necessarily to be implemented.

This kind of filter is scalable, i.e. its basic unit stage can be supplemented by any number
of the same basic unit. The adjustable parameters of such a filter type are
the gate width (each stage is individually configurable) and the number of consecutive
filter stages. When building up multiple filter stages in series the following choice
of the time constants is of practical interest:
${\rm gate(1)}\le{\rm gate(2)}=\dots={\rm gate(N)} \equiv {\rm gate(2 \dots N)}$.

Choosing the time constants for all the stages equal means that the initial timing criterium
defined by the parameter gate(1) has to be fulfilled N times.
To pass N stages of such a multistage filter the SD pulses
at its input should appear in groups of at least ${\rm N}+1$ pulses
with temporal distances between them not exceeding the value of gate(1).
The first N pulses of such a group will be removed by the filter and
the rest will pass through.
The residual dark-count rate $n$ after stage N of the filter can be approximated
with the following empirical dependence:

\begin{equation}
n = n_0\ (1 - \alpha)\ \alpha^{\rm N}, \qquad \alpha = \Delta\ n_0
\label{Eq:n=f(N)}
\end{equation}
where $n_0$ is the dark count rate at the filter input and
$\Delta$ is the coincidence time at the first filter stage calculated as the sum
of gate(1) and the width of the SD pulse ($\sim 10$\,ns, see Figure~3).

Selecting the time constants ${\rm gate(2 \dots N)}$ greater than gate(1)
allows for gaps up to this new longer time constant within
the pulse sequence decimated by the first stage.
This choice affects only slightly the dark-count rejection capabilities of the filter,
minor increase of the parameter $\alpha$ in (\ref{Eq:n=f(N)}),
but it is definitely more favorable for
transmitting the scintillation signals as they are characterized by an increased SD pulse
density over a rather long (microseconds) period of time.

Figure~5 shows an example of measured SD and SF pulse sequences from the detection
of one neutron scintillation event. The first pulse of the SF signal
(a subsequence of SD pulses passing the filter) is triggering the event generator
(a retriggerable mono-flop with adjustable pulse width) generating an event signal SN.
The width of the SN signal is set to prevent multiple triggers from the same scintillation event
(late after-glow photon elimination).
Note, that the appearance of a second neutron event within the duration of the first one
will lead to an increase of the density (and not the amplitude) of the analog signals
SA after the amplifier and to a corresponding increase of the density
of the SD pulses after the discriminator.
These two events will be detected as one event
if the time interval between them is smaller compared to the blocking time of the event generator
(depending on the strength of the second event the blocking time might be prolonged)
or as two events otherwise.

\section{Measurements}
The trigger efficiency of the detector was measured using a calibrated alpha source.
To find out how representative the measurements with the alpha source are,
the distribution of the number of detected photons measured with alpha particles was compared
with that measured with neutrons from a weak non-calibrated neutron source.

The $^{241}$Am alpha source is a tablet with a diameter of $\oslash = 25$\,mm
covered with a 0.2\,mm thick aluminium plate having a 16\,x\,1.6\,mm slit serving
as a collimator. The source was positioned below the scintillator structure (see Figure~1).
The energy of the alpha-particles is $\sim 5.5$\,MeV, their penetration depth into
the scintillator is $\sim 20$\,$\mu$m.
Consequently, the minimum path for the scintillation light from the point of emission
to absorption in a WLS fiber is $\sim 130$\,$\mu$m. For absolute efficiency measurements
the source was calibrated in the following way: a plastic scintillator was mounted
on the photocathode of a photomultiplier tube and the intensity of alpha-particles
was measured to be $94\,\pm\,1$~1/s. The detection threshold was set low enough to ensure
that all alpha-events were detected.

Thermalized neutrons were obtained from a $^{241}$AmBe source
(activity $\sim 2\cdot10^4$~neutrons/s) enclosed in a polyethylene moderator.
The rate of thermal neutrons detected with our test detector unit was in the order
of 1 count per second. As the neutron interactions are distributed more uniformly
within the scintillator volume compared to the alpha-particle interactions close
to the surface of the scintillator, the light collection conditions are expected
to be more favorable in case of neutrons.

The performance of the detector is fully determined by the following
parameters of the signal processing system:
discrimination threshold on the SiPM analog signals,
the number of filtering stages N,
the gate widths of the first gate(1) and the following ${\rm gate(2 \dots N)}$ filter stages,
and the duration of the output pulse of the event generator (blocking time).

The discrimination threshold can be set at any value below the minimum amplitude
of the SiPM one-cell signals (see Figure 3):
as long as all these signals are accepted the threshold value
does not influence the parameters of the detector.

The blocking time was fixed at a rather large value of 100~$\mu$s to ensure that in all range
of variation of the filter parameters the multicount ratio is below 1\,\%.

The filter gate widths ${\rm gate(2 \dots N)}$ were fixed at the value of 200~ns to ensure
the fulfillment of the condition ${\rm gate(1)} \le {\rm gate(2 \dots N)}$ in the whole range
of the foreseen variation of the parameter gate(1).

The range of variation of the parameters gate(1) and N was established from
the following consideration.
An efficient suppression of the SiPM dark counts with
a reasonable (about 10) number of filter stages requires,
see formula (\ref{Eq:n=f(N)}), the parameter $\alpha$ to be in the order of 0.1 or below.
At the dark count rate of 1\,MHz this corresponds to ${\rm gate(1)} \le 100$\,ns.
Suppression of the dark count rate to at least $10^{-2}$\,Hz, i.e. by 8 orders of magnitude,
would require with gate(1)\,=\,100\,ns about ${\rm N} = 8$ filter stages.
Accordingly, the parameters gate(1) and N were varied in the present measurements
as: ${\rm gate(1)} = 50 - 140$\,ns and ${\rm N} = 8, 10$.

Figure~6 shows the measured trigger efficiency for alpha-particles as a function
of the parameter gate(1) at N\,=\,8 and 10.
The required efficiency of $\ge 80$\,\% is achieved at N\,=\,8 and
$\rm{gate(1)} \ge 80$\,ns.

Figure~7 shows the rate of residual dark counts at the filter output as a function
of the total number N of filtering stages at gate(1)\,=\,80\,ns.
Each stage is suppressing the dark-count rate
by a factor of about 10, so that at $\rm{N} = 8$
(where we get the trigger efficiency of $\sim 82$\,\%)
the residual output dark-count rate (background count rate of the detector)
reaches the required value of about $3\cdot10^{-3}$\,Hz.

Figure~8 shows the distributions of accumulated numbers of SD-pulses corresponding
to the detection of alpha-particles and neutrons at gate(1)\,=\,80\,ns and N\,=\,8.
The SD-signal count accumulation time
was set to the first 10\,$\mu$s of the detected events.
This parameter (number of SD-counts) was obtained using the measuring capabilities
of a LeCroy WaveRunner 640Zi DSO. In case of neutron detection the average number
of photon counts is about 1.5 times larger.
This fact allows us to consider the trigger efficiency determined above
in measurements with the alpha source as a conservative estimate for the trigger efficiency
with neutrons.

\section*{Summary}
The feasibility of using Silicon Photomultipliers operated at room temperature
for individual channel readout
in multi-channel neutron detection systems utilizing ZnS:$^6$LiF scintillation screens
is demonstrated. An efficient light collection is achieved by uniformly distributing
$\oslash = 0.25$\,mm WLS fibers within the volume of the scintillator.
A signal processing scheme based on the acceptance of all
of the SiPM pulses (generated either by scintillation photons or thermally)
with a subsequent efficient suppression of the dark counts is presented.

With a prototype singe-channel detection unit using a SiPM with dark-count rate of $\sim 1$\,MHz
the background count rate of $\sim 3\cdot10^{-3}$\,Hz
and the trigger efficiency for alpha-particles of $\sim 80$\,\% are demonstrated.
For neutrons a higher trigger efficiency
is to be expected due to the larger amount of photons from one single neutron absorption process.

Optimization of the design of the scintillator\,/\,WLS fiber structure
(type of used fibers, fiber spacing both within one layer and between the layers)
as well as a full characterization of the detector with a neutron source
will be subjects of further investigations and discussed elsewhere.

\section*{Acknowledgments}
We express our gratitude to Andreas Hofer (Detector Group of the Laboratory for Particle Physics)
for designing and building our prototype detection units.


\clearpage
\newpage

\begin{figure}[t]
\centering
\includegraphics[width=0.7\columnwidth,clip]{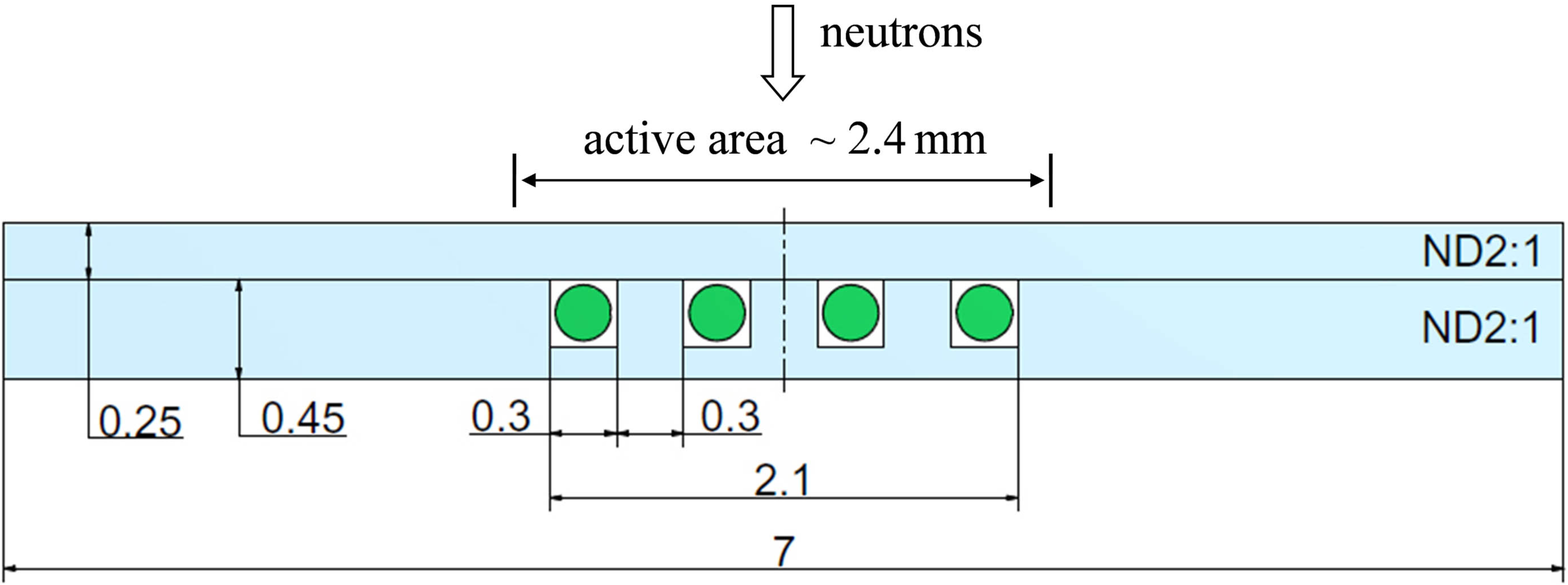}
\caption{
Test scintillator unit used in the present measurements.
The unit consists of two layers of ZnS:$^6$LiF scintillation material
with WLS fibers embedded into the grooves machined in the thicker layer.
The effective channel width is defined by the area covered by the fibers and is about 2.4\,mm.
The additional space to the total width of 7\,mm is used for handling purposes only.
}
\end{figure}

\begin{figure}[b]
\centering
\includegraphics[width=1.0\columnwidth,clip]{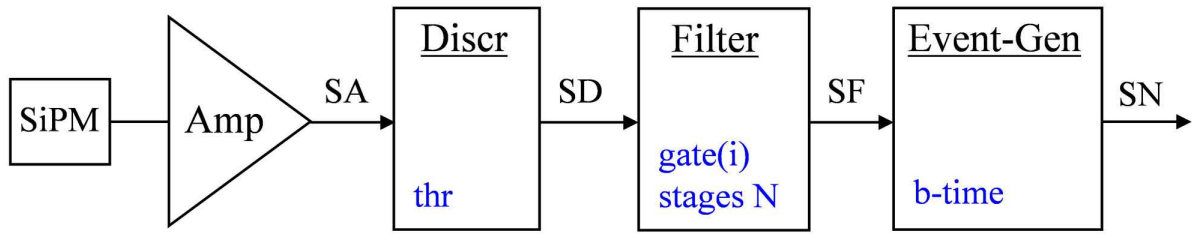}\\[2ex]
\caption{
Processing scheme of the SiPM signals: high band-width amplifier,
leading-edge discriminator stage, multistage filter, and event generator unit.
The tunable parameters are: the discrimination threshold (thr),
the number of filter stages (N),
the retriggerable gate width of each of the filter stages gate(i), and
the duration of the output pulse of the event generator (b-time - blocking time)
to account for the after-glow photons from the scintillation process.
}
\end{figure}

\clearpage
\newpage
\begin{figure}[t]
\centering
\includegraphics[width=0.7\columnwidth,clip]{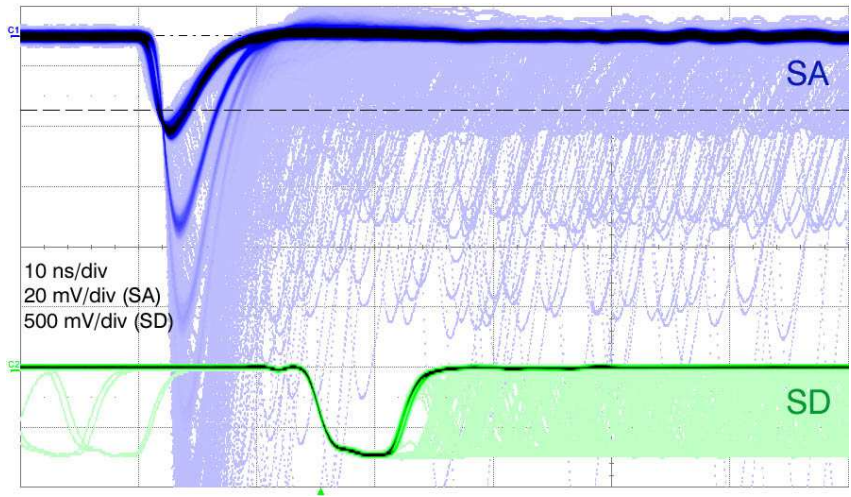}
\caption{
Conversion of the amplified and shaped SiPM analog signals (SA)
into standard digital signals (SD) by a fast leading edge discrimination stage.
Note, that independent of how many SiPM cells fire simultaneously (via optical cross-talk)
per single primary charge carrier initiating the avalanche in one of the cells
(see multiple amplitudes of SA signals),
after the discriminator we always have the correspondence
``one primary charge carrier - one standard SD pulse'',
i.e. in contrast to the SA signals, the SD signals are free from the cross-talk contribution.
}
\end{figure}

\begin{figure}[b]
\centering
\includegraphics[width=1.0\columnwidth,clip]{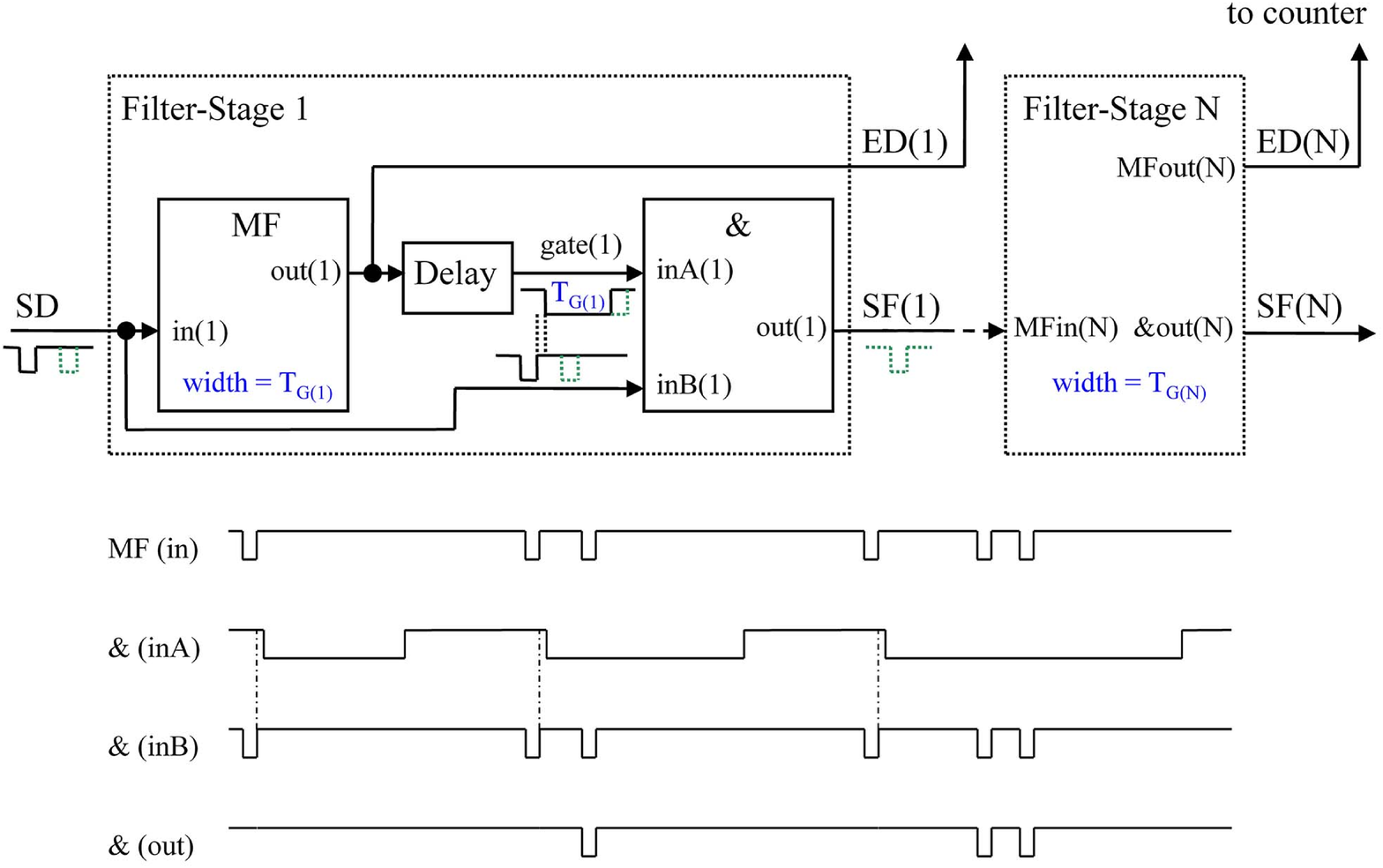}
\caption{
Block diagram (top) of a ``single-pulse elimination'' multistage filter:
MF~-~retriggerable mono-flop, \&~-~coincidence logic.
The filter performs a consecutive delayed self-coincidence signal processing
on the input SD-pulse sequence. The ED(i) outputs enable to measure the residual count rate,
whereby the rate after the stage~1 is measured at ED(2), after the stage~2 at ED(3) and so on.
The timing diagram (bottom) illustrates the passing of pulses through a filtering stage.
For clarity the internal delays of MF and \& units are assumed to be zero.
The active state is low due to the standard NIM-logic used.
}
\end{figure}

\clearpage
\newpage
\begin{figure}[t]
\centering
\includegraphics[width=1.0\columnwidth,clip]{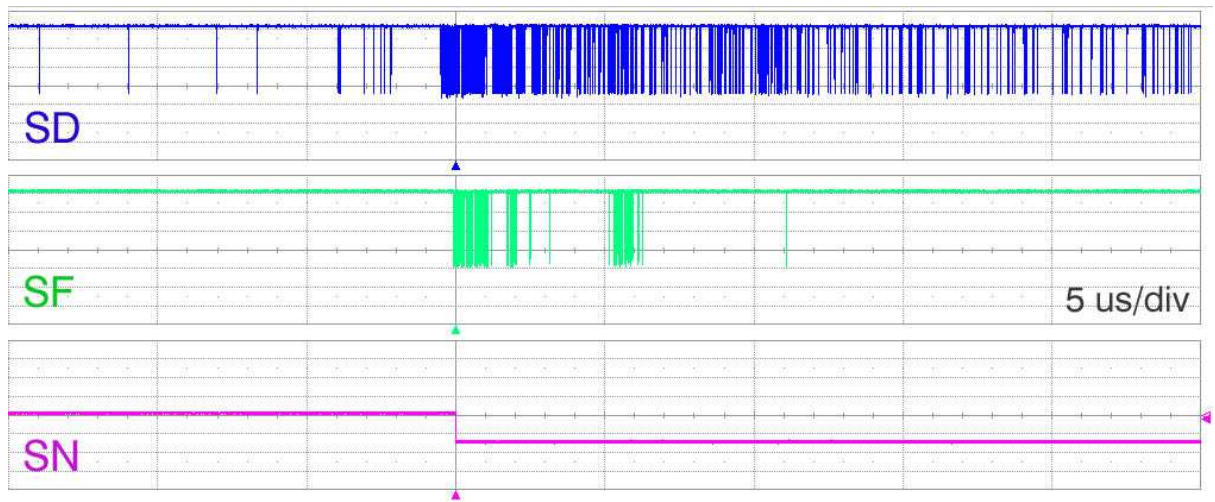}
\caption{
Detection of a neutron event: initial SD pulse sequence and the subsequence of the SD-pulses
passing the filter (SF-signal).
The filter settings are: gate(1)\,=\,80\,ns, ${\rm gate(2 \dots N)} = 200$~ns, N\,=\,8.
The first pulse of the SF-sequence is generating
the leading edge of an event signal (SN). The width of the SN-signal
(retriggerable mono-flop type) was set in the present measurements at 100\,$\mu$s.
}
\end{figure}

\clearpage
\newpage
\begin{figure}[t]
\centering
\includegraphics[width=0.7\columnwidth,clip]{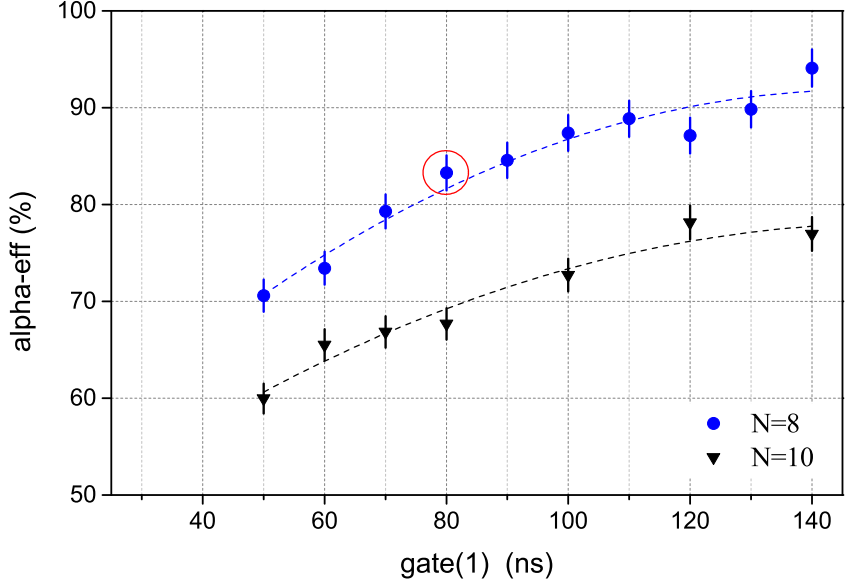}
\caption{
Alpha-particle detection (trigger) efficiency as a function of the gate width
of the first filter stage (all other filter gates are set to 200\,ns).
The total number of filter stages is fixed to N~=~8 or 10.
The collimating slit of the alpha-source is oriented along the axis
of the scintillator structure.
The selected filter setting for further measurements is indicated by the red circle.
The dashed lines are shown to guide the eye.
}
\end{figure}

\begin{figure}[b]
\centering
\includegraphics[width=0.7\columnwidth,clip]{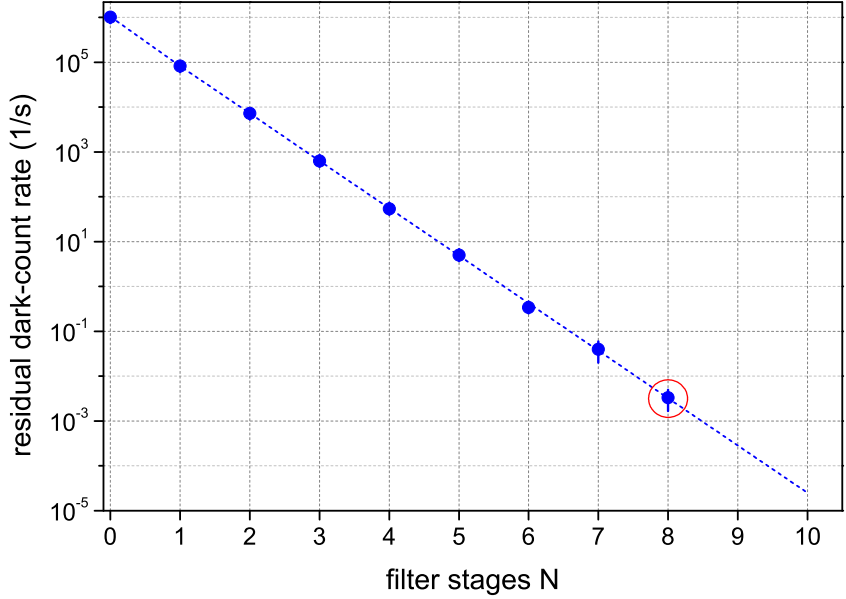}
\caption{
Residual dark-count rate at the filter output as a function of the number
of filter stages N. Dark-count bunches (groups) with temporal distances between
individual pulses of $\le$\,gate(i) width are just counted as one single pulse.
The gate-widths are fixed to: gate(1)\,=\,80\,ns, ${\rm gate(2 \dots N)} = 200$~ns.
The dashed line is shown to guide the eye:
starting from N\,=\,1 it corresponds to a fit using the function (\ref{Eq:n=f(N)})
with the parameters $n_0 = 1040$\,kHz and $\alpha = 0.088$.
The point at N\,=\,0 denotes the measured value for the input dark count rate $n_0$.
The selected filter setting for further measurements is indicated by the red circle.
}
\end{figure}

\begin{figure}[b]
\centering
\includegraphics[width=0.85\columnwidth,clip]{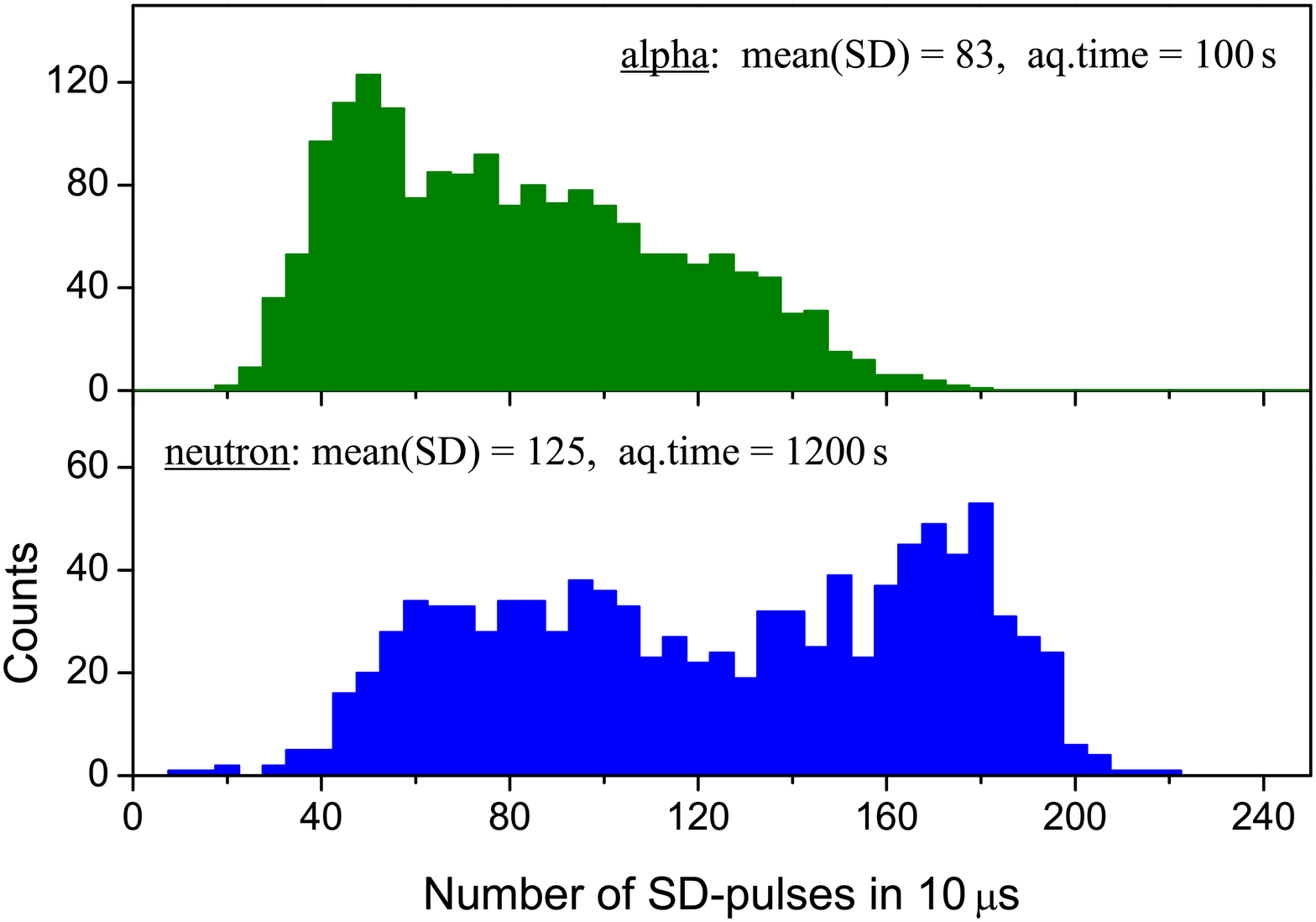}
\caption{
Distributions of the accumulated numbers of SD-counts for the detection of alpha-particles
and thermalized neutrons within the first 10\,$\mu$s of the corresponding events
measured at gate(1)\,=\,80\,ns, $\rm{gate(2 \dots N)} = 200$\,ns, and N\,=\,8.
In both cases no collimators were used, accordingly some of the detected events might
come from interactions outside the region of optimum light collection covered by the WLS fibers.
}
\end{figure}


\begin{thebibliography}{99}

\bibitem{Rhodes97} N.J.~Rhodes et al., Nucl.~Instr.~and~Meth.~A 392 (1997) 315. \vv

\bibitem{Sakasai09} K.~Sakasai et al., Nucl.~Instr.~and~Meth.~A 600 (2009) 157. \vv

\bibitem{Nakamura12} T.~Nakamura et al., Nucl.~Instr.~and~Meth.~A 686 (2012) 64. \vv

\bibitem{Nakamura14} T.~Nakamura et al., Nucl.~Instr.~and~Meth.~A 741 (2014) 42. \vv

\bibitem{Dolgoshein06} B.~Dolgoshein et al., Nucl.~Instr.~and~Meth.~A 563 (2006) 368. \vv

\bibitem{Collazuol11} G.~Collazuol et al., Nucl.~Instr.~and~Meth.~A 628 (2011) 389. \vv

\bibitem{Kuzmin02} E.S.~Kuzmin et al., Journal of Neutron Research 10(1) (2002) 31. \vv

\bibitem{Mosset13} J.-B.~Mosset et al., arXiv:1309.6885v1 \vv

\bibitem{Hamamatsu} {\tt http://www.hamamatsu.com} \vv

\bibitem{Musienko07} Y.~Musienko et al., Nucl.~Instr.~and~Meth.~A 581 (2007) 433. \vv

\bibitem{AST} {\tt http://www.appscintech.com} \vv

\bibitem{Eljen} {\tt http://www.eljentechnology.com} \vv

\bibitem{Kuraray} {\tt http://www.kuraray.co.jp} \vv


\end{thebibliography}
\end{document}